\documentclass{emulateapj}
\usepackage{natbib,aas_macros}

\newcommand{\cs}{c_{\rm s}}

\newcommand{\rg}{r_{\rm gyro}}
\newcommand{\vect}[1]{\mbox{\boldmath${#1}$}}

\shorttitle{STOCHASTIC ACCELERATION IN MRI TURBULENCE}
\shortauthors{Kimura, Toma, Suzuki, \& Inutsuka}

\begin{document}

\title{STOCHASTIC PARTICLE ACCELERATION IN TURBULENCE GENERATED BY THE MAGNETOROTATIONAL INSTABILITY}
\author{Shigeo S. Kimura\altaffilmark{1,2}, Kenji Toma\altaffilmark{1,2}, Takeru K. Suzuki\altaffilmark{3}, and Shu-ichiro Inutsuka\altaffilmark{3}}
\affil{$^1$Frontier Research Institute for Interdisciplinary Sciences, Tohoku University, Sendai 980-8578, Japan}
\affil{$^2$Astronomical Institute, Tohoku University, Sendai 980-8578, Japan}
\email{shigeo@astr.tohoku.ac.jp}
\affil{$^3$Department of Physics, Nagoya University, Nagoya, Aichi 464-8602, Japan}

 \begin{abstract}
  We investigate stochastic particle acceleration in accretion flows. It is believed that the magnetorotational instability (MRI) generates turbulence inside accretion flows and that cosmic rays (CRs) are accelerated by the turbulence. We calculate equations of motion for CRs in the turbulent fields generated by MRI with the shearing box approximation without back reaction to the field. The results show that the CRs randomly gain or lose their energies through the interaction with the turbulent fields. The CRs diffuse in the configuration space anisotropically: The diffusion coefficient in direction of the unperturbed flow is about twenty times higher than the Bohm coefficient, while those in the other directions are only a few times higher than the Bohm. The momentum distribution is isotropic, and its evolution can be described by the diffusion equation in momentum space where the diffusion coefficient is a power-law function of the CR momentum. We show that the shear acceleration efficiently works for energetic particles. We also cautiously note that in the shearing box approximation, particles that cross the simulation box many times along the radial direction suffer unphysical runaway acceleration by the Lorentz transformation, which needs to be taken with special care. 
 \end{abstract}

\keywords{acceleration of particles --- turbulence --- accretion, accretion disks --- galaxies: nuclei}

\section{INTRODUCTION}\label{sec:intro}

When the mass accretion rate onto a black hole is sufficiently lower than the Eddington rate, a radiation inefficient accretion flow (RIAF) is formed \citep{ny94,ACK95a,OM11a,YWB12a,RTQ15a}. Inside RIAFs, the plasmas are so hot and tenuous that the non-thermal particles (cosmic rays; CRs) can naturally be generated \citep[e.g.,][]{qg99,tt12,nxs13,ktt14,NSX15a}. Generally in accretion flows, the magnetorotational instability (MRI) is believed to play a crucial role for transporting angular momentum. MRI generates strong and turbulent magnetic fields inside RIAFs, in which the CRs are likely to be accelerated through the stochastic acceleration process \citep{dml96,LPM04a}. This process has been analytically formulated through the quasi-linear theory, and the evolution of momentum distribution function of the CRs has been usually described by the diffusion equation in momentum space \citep[e.g.][]{SM98a,sp08,Ohi13a}. \citet{kmt15} investigate the stochastic acceleration inside RIAFs of low-luminosity active galactic nuclei by using this formulation and find that CRs can be accelerated up to $\sim 10$ PeV if the turbulence has the Kolmogorov spectrum, and that high energy neutrinos generated through the interaction between the CRs and background matter are compatible with the IceCube events \citep{ice13prl,ice15a}. Protons escaping from RIAFs interact with circumnuclear matter and emit gamma-rays \citep{FKM15a}, which can be consistent with the observed TeV flux from Sgr A* \citep{hess09gc} and Cen A \citep{hess09cenA}. These indicate that understanding the acceleration process in accretion flows is imperative to study the sources of such astrophysical neutrinos and gamma-rays. 
 
There are many other candidate sites of stochastic acceleration process, such as clusters of galaxies \citep[e.g.,][]{Bla00a,Pet01a,BL07a}, gamma-ray bursts \citep[e.g.,][]{AT09a,MAT12a}, and blazars \citep[e.g.,][]{KGM06a,ATK14a,KTA15a}. These studies also use the quasi-linear theory, which is constructed under the assumption of weak and isotropic turbulence, while some of those sites may include strong and/or anisotropic turbulences. Some numerical works support the validity of the quasi-linear theory even for the case with strong turbulence \cite[e.g.,][]{CLP02a,DMS03a,ort09}, but recent particle simulations show substantial differences between the simulation results and the quasi-linear theory \cite[e.g.,][]{lyn+14,FM14a,TWJ14a}. In these simulations, the turbulence are given with some Fourier spectra or driven in magnetohydrodynamic (MHD) fluids fluctuated by some algorithms. Although these simulations are useful for investigating physics of stochastic acceleration owing to controllability of the turbulence, it is also important to study particle acceleration in realistic astrophysical turbulences, which are generated by MHD instabilities, such as Kelvin-Helmholtz instability and MRI. The anisotropy of flow is essential to grow these instabilities, and MHD simulations show that non-linear growth of these instability generates turbulent magnetic fields stronger than the mean magnetic fields \citep[e.g.,][]{HGB95a,SI09a,ZMW09a}. We study the particle acceleration under the turbulence induced by such physical instability. 

Recently, particle in cell (PIC) simulations are performed to study plasma kinetics in RIAFs \citep{riq+12,hos13,RQV15a,hos15}. They show that turbulent fields are generated by MRI, and the magnetic reconnections generate CRs.  These PIC simulations need to resolve the gyro-motion of thermal particles, so that practically they have to adopt quite strong shears or weak magnetic fields even though these conditions are not realistic.

In this paper, we assume thermal particles are described by MHD, and calculate the orbital motions of test particles in the turbulent fields generated by MRI. At present, only MHD simulations can treat the large scale turbulent fields, which are essential to determine the maximum energy of accelerated particles in accretion flows, because the particles with higher energy have larger gyro radii. Our approach makes it possible to investigate the stochastic acceleration with more realistic turbulences and shear flows.  However, our treatment cannot take account of the injection process for thermal particles into non-thermal acceleration. In this sense, this study is complementary to the PIC simulations of MRI. Our method is similar to that of \citet{RII16a}, who solve the particle orbits under turbulences generated by Richtmyer-Meshkov instability in supernova remnants \citep{ISO13a}. 

This paper is organized as follows. Section \ref{sec:mhd} describes the numerical scheme and characteristics of turbulence generated in MHD simulations. The method and results of particle simulations are presented in Section \ref{sec:particle}. We discuss some features of our simulations in Section \ref{sec:discussion}, and Section \ref{sec:summary} is devoted to a summary.

\section{MAGNETO-HYDRODYNAMICAL SIMULATION}
\label{sec:mhd}

\subsection{Numerical Scheme}

First, we perform ideal MHD simulations to obtain turbulent fields generated by MRI. We ignore the back reaction by CRs for simplicity. We use the isothermal equation of state, and do not solve the energy equation. We use the shearing box approximation without vertical gravity for simplicity \citep{HGB95a}. The shearing box is a local approximation of differentially rotating plasma. We use the co-rotational frame and ignore the curvature effect. Then, the set of MHD equations is written as 
\begin{eqnarray}
&& {\partial \rho \over \partial T} + \nabla \cdot \left(\rho \vect V_{\rm fl}\right) = 0, \\
&& {\partial \vect B \over \partial T}  = - \nabla \times \left(\vect V_{\rm fl} \times \vect B\right), \\
  && {\partial \vect V_{\rm fl} \over \partial T} + \vect V_{\rm fl} \cdot \nabla \vect V_{\rm fl} = - {1 \over \rho} \left( P + {B^2\over 8\pi}\right) \nonumber \\
 && ~~~~~ + {(\vect B\cdot \nabla)\vect B \over4\pi \rho} - 2\vect \Omega_{\rm K} \times \vect V_{\rm fl} + 3 \Omega_{\rm K}^2 x \vect e_x,
\end{eqnarray}
where $\vect e_i$ is the unit vector of $x_i$ direction, $\vect V_{\rm fl}$ is the fluid velocity, $\vect \Omega_{\rm K}=\Omega_{\rm K}\vect e_z$ is the angular velocity at box center ($\Omega_{\rm K}$ is the Keplerian angular velocity), $T$ is the time for the MHD simulation, and the other letters have usual meanings. In the shearing box, the direction $x,~y,$ and $z$ correspond to $r,~\phi$, and $z$ in the cylindrical coordinate, respectively. 
In this paper, we ignore the effect of resistivity, which may affect the particle acceleration through the generation of electric fields (see Section \ref{sec:plasma}).

The boundary conditions of shearing box for the magnetic fields are described as
\begin{eqnarray}
\vect B(x,y,z)&=&\vect B(x+L_x,~y-1.5\Omega_{\rm K} L_x T,~z), \label{eq:bc_shear} \label{eq:bc_x}\\
\vect B(x,y,z)&=&\vect B(x,~y+L_y,~z)\label{eq:bc_y}, \\ 
\vect B(x,y,z)&=&\vect B(x,~y,~z+L_z)\label{eq:bc_z}.
\end{eqnarray}
The same conditions are used for the velocity fields and density. There is the background shear velocity in a shearing box, $\vect V_{\rm shear}=-1.5\Omega_{\rm K} x \vect e_y$, where the factor 1.5 comes from the index of radial dependence of angular velocity, $\Omega_{\rm K} \propto r^{-1.5}$. This generates relative velocities in the $y$ direction between the boxes adjacent in the $x$ direction, so that the $y$ coordinate in Equation (\ref{eq:bc_shear}) depends on the MHD time $T$. As the initial condition for MHD simulations, we set the plasma of uniform density $\rho_0$. The initial plasma has a vertical magnetic field, $B_0 = (0,~0,~B_z)$, with $\beta_{\rm pl}=10^4$, where $\beta_{\rm pl} = 8\pi \rho_0 c_{\rm s}^2/B_0^2$ ($c_{\rm s}$ is the sound speed). We perform MHD simulations with the above setup using a modified version of a MHD code used for investigating MRI turbulence in protoplanetary disks \citep{SI09a,SMI10a}, which adopts a second-order Godunov-CMoCCT scheme.

The most unstable wave length of MRI, $\lambda_{\rm cr}$, is represented as  \cite[e.g.,][]{BH98a,san+04}
\begin{equation}
 \lambda_{\rm cr} = {8\pi \over \sqrt{15}}{v_{\rm A,0} \over \Omega_{\rm K}} 
\simeq 0.0649 H \left(\frac{\beta_{\rm pl}}{10^4}\right)^{-1/2}, \label{eq:lambda_cr}
\end{equation}
where $v_{\rm A,0}=B_0/\sqrt{4\pi \rho_0}$ is the Alfv\'{e}n velocity, and $H=\sqrt 2 \cs/\Omega_{\rm K}$ is the scale height. For the particle simulation, the box size and mesh number are fixed as $(L_x/N_x,~L_y/N_y,~L_z/N_z) = (2H/128,~4H/256,~2H/128)$, for which $\lambda_{\rm cr}$  is resolved by several grids. We briefly discuss the effect of the box size in the next subsection.

\subsection{Characteristics of MRI Turbulence}

\begin{table}[tb]
\begin{center}
\caption{mean values of background fields \label{tab:mean_values}}
\begin{tabular}{|c|c|}
\hline
$\langle B_x^2\rangle_{\rm v} /(8\pi P_0)$ & $ 6.58\times10^{-3}$\\
$\langle B_y^2\rangle_{\rm v} /(8\pi P_0)$ &  $3.62\times10^{-2}$ \\
$\langle B_z^2\rangle_{\rm v} /(8\pi P_0)$  & $3.23\times10^{-3}$ \\
     \hline
$\langle \delta V_x^2\rangle_{\rm v} /c_{\rm s}^2$  & $1.79\times10^{-2}$ \\
$\langle \delta V_y^2\rangle_{\rm v} /c_{\rm s}^2$  & $2.66\times10^{-2}$ \\
$\langle \delta V_z^2\rangle_{\rm v} /c_{\rm s}^2$  & $1.10\times10^{-2}$ \\
     \hline
$\langle E_x^2\rangle_{\rm v} /(8\pi P_0)$  & $2.63\times10^{-5}$ \\
$\langle E_y^2\rangle_{\rm v} /(8\pi P_0)$ & $5.07\times10^{-7}$  \\
$\langle E_z^2\rangle_{\rm v} /(8\pi P_0)$  & $5.45\times10^{-5}$\\
\hline
\end{tabular}
\end{center}
\end{table}

  \begin{figure}[tbp]
   \begin{center}
    \includegraphics[width=8cm]{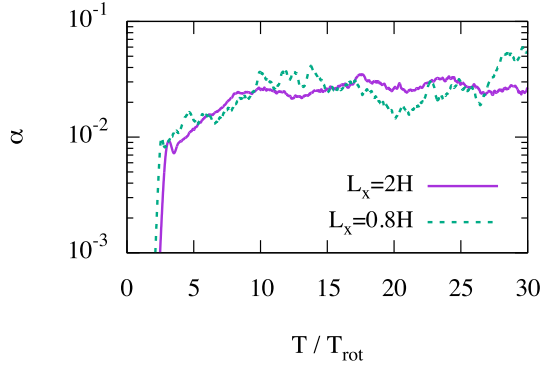}
    \caption{The evolution of $\alpha$ for the turbulent fields generated by MRI. The solid and dotted lines represent models with $(L_x,~L_y,~L_z) = (2H,~4H,~2H)$ and $(0.8H,~1.6H,~0.8H)$, respectively. The quasi-steady state in the model with the smaller box fluctuates more strongly than that with the larger box.}
    \label{fig:alpha}
   \end{center}
  \end{figure}

Non-linear growth of MRI generates strongly turbulent fields. After several orbital periods, a quasi-steady state is realized.  The strength of turbulent fields are indicated by the $\alpha$ parameter introduced by \citet{ss73}, which is expressed as
\begin{equation}
 \alpha= {w_{xy} \over P_0} =  {1 \over P_0} \left\langle \rho \delta V_x \delta V_y - {B_x B_y \over 4\pi}\right\rangle_{\rm v}
\end{equation}
where $w_{xy}$ is the $xy$ component of stress tensor, $\vect{\delta V}$ is the turbulent velocity, $P_0=\rho_0 \cs^2$ is the initial pressure, and $\langle \rangle_{\rm v}$ denotes the volume average quantities.  The turbulent velocity is defined as $\vect{\delta V}=\vect V_{\rm fl} - \vect V_{\rm shear}$. Figure \ref{fig:alpha} shows the evolution of $\alpha$ for two models of different box size, $(L_x,~L_y,~L_z) = (2H,~4H,~2H)$ and $(0.8H,~1.6H,~0.8H)$. We can see that for the model with the larger box, $\alpha$ is almost constant ($\sim 0.02 - 0.03$) for $T\gtrsim7~ T_{\rm rot}$, where $T_{\rm rot}=2\pi/\Omega_{\rm K}$ is the rotation time. On the other hand, $\alpha$ varies between 0.01 and 0.06 for the smaller one. This indicates that the smaller the box size is, the more strongly the quasi-steady state fluctuates \citep{BMC08a,JSD13a}. Hereafter, we will use the data set of the larger box model in order to avoid the effect of temporal fluctuation of turbulent fields. The initial value of $\beta_{\rm pl}$ also affects the saturation level of MRI, but we only treat the data set for $\beta_{\rm pl}=10^4$ as the first step of our study.

We tabulate the components of the averaged magnetic field $\langle B_i^2 \rangle_{\rm v}$, turbulent velocity $\langle \delta V_i^2 \rangle_{\rm v}$, and electric field $\langle E_i^2 \rangle_{\rm v}$ in Table \ref{tab:mean_values}, for which we use MHD data of $T=20~T_{\rm rot}$.  The electric field is computed from the ideal MHD condition,
\begin{equation}
 \vect E = - {\vect V_{\rm fl} \times \vect B\over c}. \label{eq:Ecalc}
\end{equation}
These fields satisfy $\langle B_y^2 \rangle_{\rm v} > \langle B_x^2 \rangle_{\rm v} \gtrsim \langle B_z^2 \rangle_{\rm v}$, $\langle \delta V_y \rangle_{\rm v} \gg \langle \delta V_x \rangle_{\rm v} \gtrsim \langle \delta V_z \rangle_{\rm v}$, and $\langle E_z^2 \rangle_{\rm v} \gtrsim \langle E_x^2 \rangle_{\rm v} \gg \langle E_y^2 \rangle_{\rm v}$. The magnetic field is stretched by the shear motion, so that $B_y$ is stronger than the other directions. The shear velocity is faster than turbulent velocity, $\langle V_{\rm shear}^2\rangle_{\rm v} > \langle\delta V_i^2\rangle_{\rm v}$. The electric field is then mainly generated by the shear motion, so that $\langle E_y^2\rangle_{\rm v}$ is weaker than $\langle E_x^2\rangle_{\rm v}$ and $\langle E_z^2\rangle_{\rm v}$.

  \begin{figure}[tbp]
  \begin{center}
        \includegraphics[width=8cm]{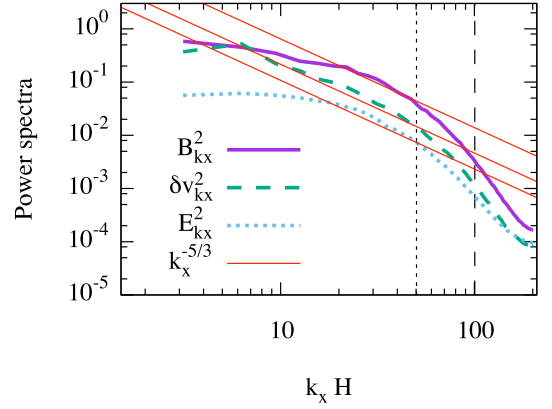}
        \includegraphics[width=8cm]{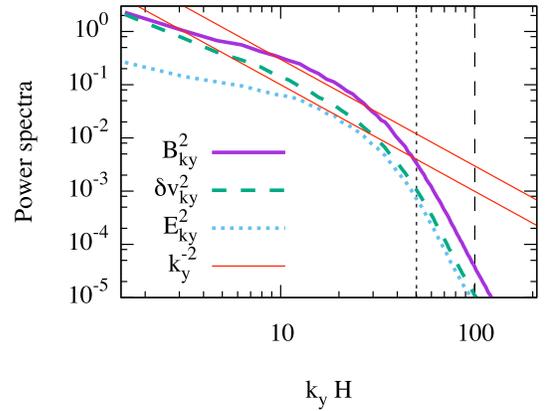}
   \caption{The power-spectra of the data of $T=20~T_{\rm rot}$. The upper and lower panels show the power spectra for $x$ and $y$ direction, respectively. The thick-solid, thick-dashed, and thick-dotted lines represent $B_k^2/(8\pi P_0)$, $\delta V_k^2/c_s^2$, and $c^2E_k^2/(8\pi P_0)$, respectively. The thin-solid lines show the spectrum predicted by \citet{gs95}, which is almost consistent with the simulation results for $10 \lesssim k H \lesssim 50$. The power-spectra rapidly decrease for $k H \gtrsim 50$ due to the numerical dissipation. The vertical dashed and dotted lines show the initial gyro-scales for model A1 and A3, respectively.   }
      \label{fig:pwsp}
    \end{center}
  \end{figure}

We calculate the power spectra of turbulent magnetic, velocity, and electric fields. First, we take the Fourier transformation of the variables 
\begin{equation}
 \vect B_{\vect k} = \int dxdydz \exp(i\vect k \cdot \vect x)\vect B(x,y,z). 
\end{equation}
Then, we define the one-dimensional power spectrum as 
\begin{equation}
 B^2_{k_x} = \int dk_y dk_z |\vect B_{\vect k}|^2,
\end{equation}
where we write down the definition of power spectrum of magnetic field for the $x$ direction. The power spectra of the other variables and the other directions are also defined in the same manner. The power spectra of $\vect B$, $\vect {\delta V}$, and $\vect E$ for the $x$ and $y$ directions of the data of $T=20~T_{\rm rot}$ are shown in Figure \ref{fig:pwsp}. The Nyquist wavenumber is $k_{\rm N}\simeq 201/H$ for all the direction, and then the small scale turbulence $k_i \gtrsim k_{\rm N}/4\simeq 50/H$, shown by the vertical dotted line, suffers from the numerical dissipation. On the other hand, the large scale turbulence for $k_i \lesssim 10/H$ is probably affected by the injection process of turbulence. This scale becomes larger as MRI grows in the non-linear regime, although it is given by Equation (\ref{eq:lambda_cr}) in the linear regime, whose scale is close to the vertical dashed line in Figure \ref{fig:pwsp}  \citep[e.g.,][]{SI01a}. Here, we consider $10 \lesssim k_i H \lesssim 50$ as the inertial range. The shapes of the power spectra in the inertial range are similar for all the variables. The power spectra have anisotropy. The spectra for the $x$ and $z$ directions are almost the same, which is almost consistent with the Kolmogorov turbulence, $B^2_{k_x}\sim B^2_{k_z}\propto k_x^{-5/3}$. On the other hand, $B^2_{k_y}\propto k_y^{-2}$, which is steeper than those for the other directions. Since the magnetic field is almost parallel to the $y$ direction, these properties are consistent with the theory of incompressible MHD turbulence \citep{gs95}, which is also confirmed by the previous simulations \citep[e.g.,][]{CL03a}. These properties of the turbulent fields are unchanged even for snapshot data of different MHD times. Note that in order to see the waves moving with the background shear, one should change the coordinates and wavenumbers according to \citet{HGB95a}. Although this slightly modifies the inertial range of the power spectra, we do not have to deal with it because the motions of the test particles are calculated in the coordinates of the shearing box (see Section \ref{sec:numerical}).

\section{TEST PARTICLE SIMULATION}
\label{sec:particle}

\subsection{Numerical Procedure}\label{sec:numerical}

We calculate the orbits of CRs as test particles in the turbulence generated in the MHD simulations. The equation of motion for a test particle is 
\begin{equation}
 {d \vect p \over dt} = e \left(\vect E_{\rm mod}+{ \vect v \times\vect B_{\rm mod}\over c} \right),\label{eq:eom}
\end{equation}
where $\vect p = \gamma m \vect v $ and $\vect v$ are the momentum and velocity of the test particle, respectively ($\gamma$ is the Lorentz factor of the particle).
The tidal and Coriolis forces are included in $E_{\rm mod}$ and $B_{\rm mod}$, respectively,  \citep{hos13}
\begin{equation}
 \vect{ E_{\rm mod}}=\vect E + {3 \gamma m \over e} \Omega_{\rm K}^2 x \vect e_x,
\end{equation}
\begin{equation}
  \vect{ B_{\rm mod}}=\vect B + {2\gamma m \over e} \vect \Omega_{\rm K}.
\end{equation}
Note that the tidal and Coriolis forces are much weaker than the electromagnetic force, and they little affect the orbits of CRs in our simulations. We assume CRs as protons and neglect the radiative and Coulomb losses. We use the Boris method \citep[e.g.,][]{BL91a} to integrate this equation, which is often used in PIC simulations.

Since the MHD data are defined only on the grid points, it is necessary to interpolate the data to the particle position. The MHD data consist of $\vect B$ and $\vect V_{\rm fl}$. First, we compute $\vect B$ and $\vect V_{\rm fl}$ at the particle position by linear interpolation. Then, we calculate $\vect E$ according to equation (\ref{eq:Ecalc}). This procedure guarantees $\vect E \cdot \vect B = 0$, which is necessary to avoid the artificial acceleration due to the interpolation \citep[cf.,][]{RII16a}. In reality, $\vect B$ and $\vect E$ evolve on MHD timescale. However, as the first step, we use the snapshot MHD data for simplicity. 

As the initial conditions, the particle positions and velocities are given such that a uniform and isotropic distribution is realized in the calculation box (hereafter, we call this box the initial box). We set monoenergetic CRs whose gyro radii $r_{\rm gyro,0}$ satisfy
\begin{equation}
 r_{\rm gyro,0}\equiv {E_{p,0} \over e\langle B\rangle_{\rm v}} =\epsilon_{\rm gyro} \Delta x,\label{eq:init_ene}
\end{equation}
where $E_{p,0}$ is the initial energy of CRs, $\epsilon_{\rm gyro}$ is a parameter of the initial energy of CRs, and $\Delta x = L_x/N_x$ is the mesh size. The initial Lorentz factor is then represented as 
\begin{equation}
 \gamma_0 = {e \langle B \rangle_{\rm v}  \epsilon_{\rm gyro} \over m c^2}{L_x\over N_x}.
\end{equation}
The value of $\gamma_0$ depends on the models of accretion flows. We tabulate $\gamma_0$ in Table \ref{tab:models}, supposing the accretion flow has $c_{\rm s}\simeq2.2\times10^9 \rm cm ~s^{-1}$, $\Omega_{\rm K}\simeq4.4\times 10^{-6} \rm s^{-1}$, and $n_0=\rho_0/m\simeq9.0\times10^7 \rm cm^{-3}$. These values correspond to the self-similar solution for RIAFs \citep{YN14a} with $m=10^8$ (Black hole mass in unit of $M_{\odot}$), $\dot m_{\rm BH} = 10^{-3}$ (mass accretion rate in unit of the Eddington ratio), $r=30$ (disk radius in unit of the Schwarzschild radius), $s=0$ (parameter of mass loss rate), and $\alpha=0.03$ \footnote{The definition of scale height in our paper, $H=\sqrt 2 c_{\rm s}/\Omega_{\rm K}$, is different from that in \citet{YN14a}, $H_{\rm YN}=c_{\rm s}/\Omega_{\rm K}$, so that the value of density is slightly modified. In addition, we assume $\Omega=\Omega_{\rm K}$ in our paper, although it is not good approximation for RIAFs (see Section \ref{sec:plasma}). }. In reality, energies of these CRs are so high that they probably escape from the flow in the vertical direction. It is desirable to calculate CRs of much lower energies, such as $\gamma_0\sim 10-100$. However, gyro-scale of such low energy CRs are too small compared to the grid scale of the MHD simulations. These particles do not change their energies significantly in the interactions with the large scale turbulence that is likely to determine the maximum energy of CRs in accretion flows. To investigate the interaction between CRs and the large scale turbulence, we calculate the orbital motions of the high energy CRs by prohibiting them from escaping to the vertical direction. Since these CRs are ultra-relativistic, we can write $E_{p,0}=p_0 c$, where $p_0$ is the initial momentum of CRs.

Owing to Equations (\ref{eq:bc_x}) - (\ref{eq:bc_z}), we can continue to calculate the orbits of CRs outside the initial box. Because of the relative velocities between boxes, we should convert the energies and momenta of CRs crossing the box boundary in the $x$ direction by Lorentz transformation as
\begin{eqnarray}
E_p'=\Gamma(E_p - \beta_{\rm box} c p_y),\label{eq:lt_e}\\
p_y'=\Gamma \left(p_y - \beta_{\rm box}{E_p\over c}\right),\label{eq:lt_p}\\
\ p_x'=p_x,\ p_z'=p_z,
\end{eqnarray}
where $\beta_{\rm box}= \mp 1.5 \Omega_{\rm K} L_x/c$ is the relative velocity between boxes (we use $-$ sign for the CRs crossing $x=+L_x/2$ boundary, and vice versa) and $\Gamma=(1-\beta_{\rm box}^2)^{-1/2}$. These $\Gamma$ and $\beta_{\rm Box}$ depend on $c_{\rm s}$ and $L_x/H$. We get $\beta_{\rm box}\simeq 0.306$ and $\Gamma-1\simeq 5.01\times10^{-2}$ for $c_{\rm s}\simeq 2.2\times 10^9$ cm s$^{-1}$ and $L_x/H=2$. Since $\vect v = c^2\vect p/E_p$, $v_x$ and $v_z$ are also changed by the transformation. We solve equation (\ref{eq:eom}) in the rest frame of a box where each CR is located.

This Lorentz transformation causes an artificial acceleration for sufficiently energetic CRs. This acceleration happens when the energetic CRs continuously cross several boxes without changing their directions of motions (see Section \ref{sec:runaway} for details). In order to avoid this, we remove the CRs that have crossed a boundary at $x = x_{\rm esc}$ as the escape particles. We define $x_{\rm esc} = (C_{\rm esc}+0.5)L_x$, where $C_{\rm esc}$ is the escape parameter, and set $C_{\rm esc}=2$ such that the relative Lorentz factor between the escape boundaries at $x = -x_{\rm esc}$ and $x_{\rm esc}$ is less than two. Although the Lorentz contraction may also affect the length scale that CRs travel, we ignore its effect for simplicity. 

We note that the boundary condition for CRs described above is different from that for the MHD simulations. The latter use the Galilean transformation, since the fluid velocity is non-relativistic, $\Gamma-1 \ll 1$. We cannot use the Galilean transformation as the boundary condition for CRs because the velocity of CRs after the Galilean transformation could exceed the speed of light, $v_y' = v_y + 1.5 \Omega_{\rm K} L_x \sim c +  1.5 \Omega_{\rm K} L_x > c$.

We set the time step as $\Delta t ={\rm min}(\Delta t_{\rm gyro},~\Delta t_{\rm cell})$, where $\Delta t_{\rm gyro}= C_{\rm safe}E_{p,0}/(ceB_{\rm max})$ and $\Delta t_{\rm cell} = C_{\rm safe} \Delta x/c $. Note that the velocities of CRs are always almost the speed of light because we focus on the ultra-relativistic particles. We use the maximum value of magnetic field in the box $B_{\rm max}$ to estimate $\Delta t_{\rm gyro}$, and set $C_{\rm safe}=0.01$.

\subsection{Results}\label{sec:results}

  \begin{table*}[tb]
   \begin{center}
    \caption{Model parameters \& physical quantities}
    \label{tab:models}
    \begin{tabular}{|c|ccc|ccc|ccc|ccc|}
     \hline
     model  & $T$ & $\epsilon_{\rm gyro}$ & $C_{\rm esc}$ & $\gamma_0$ & $t_{\rm end}$\footnote{in unit of $t_{\rm gyro,0}$} & $\delta t^{\rm\ a}$ & $D_x$\footnote{in unit of $D_{\rm Bohm}$} & $D_y^{\rm \ b}$ & $D_z^{\rm \ b}$& $q$ & $D_0$\footnote{in unit of $\overline D_p$} & $A$  \\
     \hline
     A1& $20T_{\rm rot}$ & 4 & 2 & $ 3.4\times 10^8$& 416 & $50$ & 2.3 & 17 & 1.6 & 2.38 & $1.59\times10^{-4}$ & 0.30 \\
     A2& $20T_{\rm rot}$ & 1 & 2 & $8.5\times 10^7$& 5628 &$200$ & 2.4 & 24 & 1.6 & 1.91 & $3.17\times10^{-5}$ & 0.25 \\
     A3& $20T_{\rm rot}$ & 8 & 2 & $6.8\times 10^8$& 94 & $40$ & 2.0 & 16 & 2.0 & 2.79 & $5.38\times10^{-4}$ & 0.25 \\
     B1& $15T_{\rm rot}$ & 4 & 2 & $ 3.5\times 10^8$& 447 & $50$& 2.2 & 19 & 1.5  &2.38 & $1.50\times10^{-4}$ & 0.31  \\
     C1& $25T_{\rm rot}$ & 4 & 2 &$ 3.7\times 10^8$& 438 & $50$  & 2.2 & 17 & 1.6 &2.46 & $1.45\times10^{-4}$ & 0.28 \\
     \hline
     X1& $20T_{\rm rot}$ & 4  & $\infty$ & $ 3.4\times 10^8$& $\infty$ & $200$& 3.1 & 19 & 2.0  & 0.969 & $5.83\times10^{-5}$ & 0.54 \\
     X2& $20T_{\rm rot}$ & 1 & $\infty$ & $8.5\times 10^7$& $\infty$& $200$ & 2.5 & 20 & 1.5 & 1.31 & $2.88\times10^{-5}$ & 0.46 \\
     X3& $20T_{\rm rot}$ & 8& $\infty$ & $6.8\times 10^8$& $\infty$ & $200$ & 3.5 & 21 & 3.0 & 0.654 & $6.44\times10^{-5}$ & 0.58 \\
      \hline
    \end{tabular}
   \end{center}
  \end{table*}

We show the results of the simulations and discuss the evolution of distribution function. The parameter sets are tabulated in Table \ref{tab:models}. The letters A, B, and C represent the different times $T$ of snapshot data, and we use the numbers 1, 2, and 3 to distinguish the initial energy $\epsilon_{\rm gyro}$. Group X will be discussed in Section \ref{sec:dp}. We use $N_p=2^{15} = 32768$ CR particles, and calculate their orbits until half of CRs escape from the system, the times of which are tabulated in Table \ref{tab:models} in unit of the initial gyro period, $t_{\rm gyro,0}= 2\pi r_{\rm gyro,0}/c$. For all the models, the CRs randomly gain or lose their energies through the interaction with turbulent fields, and diffuse in both the configuration and momentum spaces.

\subsubsection{Lab Frame, Box Frame, and Shear Frame}

  \begin{figure}[tbp]
  \begin{center}
        \includegraphics[width=8cm]{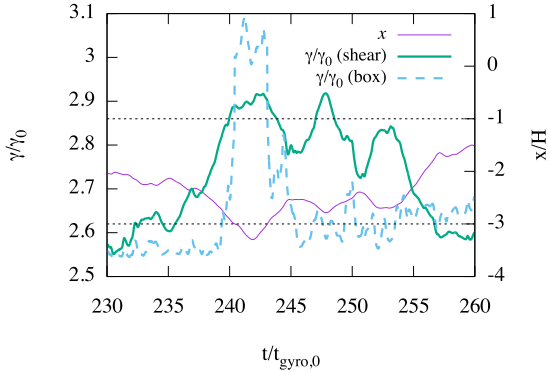}
  \caption{The evolution of energy of a CR in the shear frame (thick-solid line) and the box frame (dashed line) for model A1. The thin-solid and dotted lines show the position $x$ of the CR and the box boundary, respectively. The particle energy jumps in the box frame, while it smoothly changes in the shear frame. These jumps coincide with the CR's crossing the box boundaries in the $x$ direction. }
      \label{fig:x-gamma}
    \end{center}
  \end{figure}

There are three frames for evaluating the positions and momenta of CRs. One is the rest frame of the initial box where CRs are initially located (lab frame), another is the rest frame of a box where each CR is located at the evaluation time (box frame), and the other is the rest frame of the MHD fluid element in the mean flow (i.e., the unperturbed flow) in each box (shear frame). 

Figure \ref{fig:x-gamma} shows the evolution of the energy of a CR in the shear frame (thick-solid line) and the box frame (dashed line). When we measure the energy in the box frame, the CR energy jumps due to the Lorentz transformation. In Figure \ref{fig:x-gamma}, we can see two jumps of energy at $t\sim 240~ t_{\rm gyro,0}$ and $ t \sim 243~ t_{\rm gyro,0}$, which coincide with the CR's crossing the box boundary (shown by the dotted lines) in the $x$ direction. The CR position is represented by the thin-solid line. On the other hand, the energy measured in the shear frame does not have such jumps but smoothly evolves with time. Since the box boundaries are not special surfaces in nature, we use the shear frame for discussing the evolution of CR energy unless otherwise noted.

  \begin{figure}[tbp]
  \begin{center}
        \includegraphics[width=8cm]{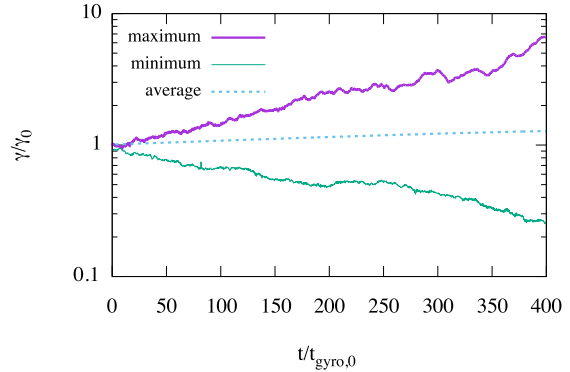}
    \caption{The long term evolution of energy of CRs in the shear frame for model A1. The thick solid line shows the evolution of energy for the most energetic CR at $t=400~ t_{\rm gyro,0}$. The thin solid line shows that for the minimum energy CR. The dotted line shows the average energy of CRs. The most energetic CR has about six times higher energy than the average value. 
    }
      \label{fig:maxmin}
    \end{center}
  \end{figure}

As mentioned above, we find that CRs randomly gain or lose a small amount of energies through interaction with the turbulent fields. A small fraction of CRs continuously gain (lose) energies, so that they can reach several times higher (lower) energies than their initial energies. Figure \ref{fig:maxmin} shows the long term evolution of the energies of such CRs in the shear frame. The most energetic CR at $t=400~ t_{\rm gyro,0}$ has about six times higher energy than the average value. This gradual change of particle energy implies that there is no ``hot spot'', where CRs gain energy efficiently, in the MRI turbulence. Note that the average energy, shown as the dotted line in Figure \ref{fig:maxmin}, is gradually increasing. This is consistent with the quasi-linear theory of the stochastic acceleration \cite[e.g.,][]{sp08}.

\subsubsection{Evolution in Configuration Space}

The diffusion coefficient in configuration space is estimated to be
\begin{equation}
 D_{x_i}= {\langle \delta x_i^2\rangle_{\rm p} \over 2\delta t}, \label{eq:diffx}
\end{equation}
where $\delta x_i = x_i(t+\delta t) - x_i(t)$ is the displacement of a particle ($x_i=x,~y$, or $~z$), $\delta t$ is the time span for estimating the diffusion coefficient, and $\langle \rangle_{\rm p}$ denotes the average over the CRs. The estimate of the diffusion coefficient is affected by the gyro motion for too short $\delta t$, while it is affected by excluding the escaped CRs for too long $\delta t$. We calculate $D_{x_i}$ for several values of $\delta t$ and choose suitable $\delta t$ for estimating the diffusion coefficient, which is tabulated in Table \ref{tab:models}. 

\begin{figure}[tbp]
  \begin{center}
        \includegraphics[width=8cm]{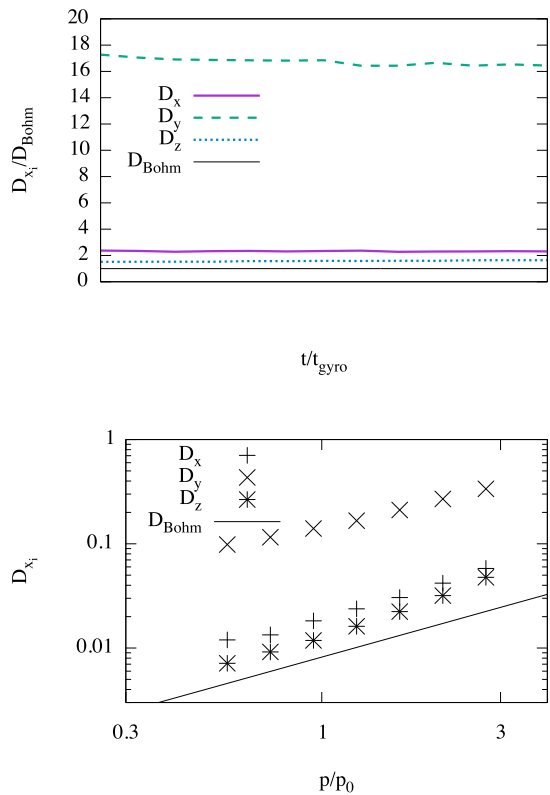}
    \caption{The diffusion coefficient in configuration space for model A1. The upper panel shows the temporal evolution of $D_x$ (thick-solid), $D_y$ (dashed), and $D_z$ (dotted) normalized by the Bohm coefficient. CRs diffuse anisotropically, $D_y > D_x \gtrsim D_z$. The lower panel shows the momentum dependence of $D_{x_i}$. The plus, cross, and asterisk points show $D_x$, $D_y$, and $D_z$, respectively. The Bohm coefficients (thin-solid lines) are also shown in both panels. The energy dependence of diffusion coefficients is almost the same with that of the Bohm one.   }
      \label{fig:diffx}
    \end{center}
  \end{figure}

We show the temporal evolution of $D_x,\ D_y,$ and $D_z$ for model A1 in the upper panel of Figure \ref{fig:diffx}. We normalize $D_{x_i}$ by the Bohm coefficient,
\begin{equation}
 D_{\rm Bohm}={1\over 3} \rg c = {\langle p \rangle _p  c^2 \over 3 e \langle B \rangle _{\rm v} }. 
\end{equation}
The diffusion coefficients in configuration space have anisotropy. Since the magnetic field is stretched in the $y$ direction, CRs easily move in the $y$ direction. Thus, $D_y > D_x \sim D_z$ is satisfied. The diffusion coefficients are almost constant in time, so the super diffusion is not observed in our simulations \citep[cf.,][]{XY13a,LY14a,RII16a}. 

We do not discuss the diffusion coefficient in the direction parallel or perpendicular to the averaged magnetic field. Although we can define the volume-averaged magnetic field $\langle \vect B \rangle_{\rm v}$, it does not represent the local (in the scale of gyro radius) direction of the magnetic field due to the strong turbulence.

Dividing CRs into some momentum bins, we estimate the momentum dependence of $D_{x_i}$. We use 20 momentum bins of equal interval in the logarithmic space for $0.1 \le p/p_0 \le 20$. To reduce the statistical fluctuation, we use only the momentum bins that have more than 100 CRs. In the lower panel of Figure \ref{fig:diffx}, the momentum dependence of $D_{x_i}$ is shown at $t=400 t_{\rm gyro,0}$. We can see that $D_{x_i}\propto p$, which is the same dependence of the Bohm coefficient. These are common features for all the models. The diffusion coefficients in configuration space for the other models are tabulated in Table \ref{tab:models}. All the models have similar values of the diffusion coefficients, $D_y \sim 20 D_{\rm Bohm}$ and $D_x\sim D_z\sim 2 D_{\rm Bohm}$.

\subsubsection{Evolution in Momentum Space}

\begin{figure}[tbp]
  \begin{center}
        \includegraphics[width=8cm]{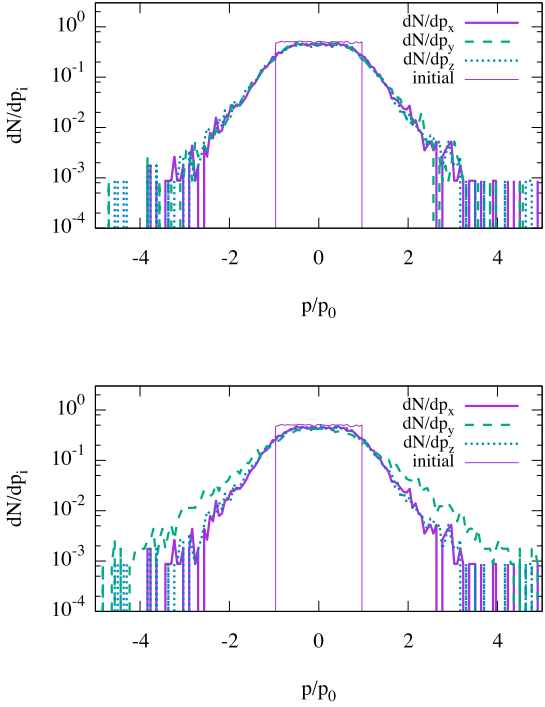}
    \caption{ The momentum distribution of each direction for model A1. The thin-solid lines show the initial distribution. The solid, dashed, and dotted lines show the final states of $dN/dp_x$, $dN/dp_y$, and $dN/dp_z$, respectively. The upper and the lower panels show the distribution in the shear frame and the lab frame, respectively. The momentum distribution is isotropic in the shear frame, while it is anisotropic in the lab frame.
    }
      \label{fig:dist_direc}
    \end{center}
  \end{figure}

We discuss the evolution of momentum distribution function $f(\vect p)$ of CRs. 
Figure \ref{fig:dist_direc} shows the momentum distribution of each direction ${dN/dp_i}$, normalized as $\int (dN/dp_i) dp_i = 1$ ($p_i=p_x,~p_y,$ or $p_z$). CRs diffuse to all the directions in momentum space. From the upper panel, it is seen that the momentum distribution is isotropic in the shear frame. The turbulent fields isotropize the motion of CRs in the shear frame. The dispersion of the momentum distribution monotonically increases with time. The momentum distribution in the box frame is almost the same as that in the shear frame because the relative shear velocity inside a box is sufficiently smaller than the particle velocities. On the other hand, in the lab frame, the distribution is anisotropic due to the velocity difference between the boxes. The dispersion in the $y$ direction is larger than the other directions. In this section, we will use the shear frame when discussing the evolution of the distribution function. Note that isotropy of the momentum distribution does not conflict with the anisotropic diffusion in configuration space. This situation is possible when the timescale of pitch angle scattering is much shorter than the diffusion timescale in configuration space. 

Since the momentum distribution is isotropic, we can write $f(\vect p) = f(p)$, where $p=|\vect p|$. Below, we show that the evolution of $f(p)$ obtained by our simulations is well described by the diffusion equation in momentum space,
\begin{equation}
 {\partial f(p)\over \partial t}
={1\over p^2}{\partial \over\partial p}
\left(p^2 D_p {\partial f(p)\over\partial p}\right) 
- {f(p)\over t_{\rm esc}}+ \dot f_{\rm inj}, \label{eq:diffusion}
\end{equation}
where $D_p$ is the diffusion coefficient in momentum space, $t_{\rm esc} = (C_{\rm esc}+0.5)^2L_x^2/(2D_x) $ is the escape time, and $\dot f_{\rm inj}=N_p \delta(t)\delta(p-p_0)$ is the injection term. We write $p = E_p/c = \gamma m c$ because CRs are ultrarelativistic. Equation (\ref{eq:diffusion}) is expected to describe the evolution of distribution function when the fractional energy change per scattering is small. 

\begin{figure}[tbp]
  \begin{center}
        \includegraphics[width=8cm]{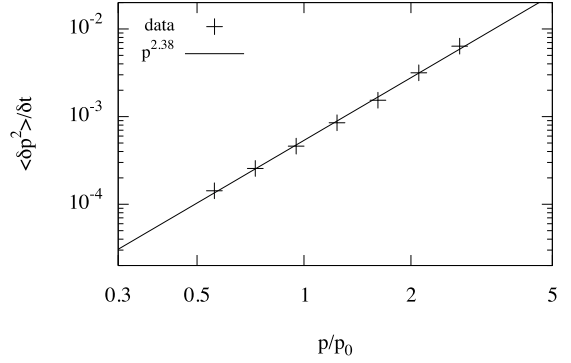}
   \caption{The momentum dependence of $\langle \delta p^2\rangle_{\rm p} / \delta t$ for model A1. The points show the simulation results, which can be fitted by a power-law function shown by the solid line.}
         \label{fig:diffp}
    \end{center}
  \end{figure}

To calculate the evolution of $f(p)$ by Equation (\ref{eq:diffusion}), we estimate the diffusion coefficient in momentum space $D_p$ from our simulation results. The diffusion coefficient is likely to be represented as
\begin{equation}
 D_p = A  {\langle \delta p^2\rangle_{\rm p}\over \delta t },
\end{equation}
where $\delta p = p(t+\delta t)-p(t)$ and $A$ is the numerical factor. We estimate $D_p$ using the momentum bins defined in the previous subsection. Although $A=1/6$ when we consider the isotropic random walk in three dimensional space, the values of $A$ are not obvious owing to the dependence of $D_p$ on $p$. In fact, the values of $A$ are different among the models (See Table \ref{tab:models}). We plot $\langle \delta p^2 \rangle_{\rm p}/\delta t $ for model A1 in Figure \ref{fig:diffp}. We can fit $ \langle \delta p^2 \rangle_{\rm p}/\delta t $ by a power-law function, $ \langle \delta p^2 \rangle_{\rm p}/\delta t  \propto (p/p_0)^q$ for all the models. The resultant $q$ is tabulated in Table \ref{tab:models}. 

We compute $A$ as follows. 
If the temporal evolution of $f(p)$ is described by Equation (\ref{eq:diffusion}), the temporal evolution of averaged momentum $\langle \dot p\rangle_{\rm p} $ is written as \citep[cf.,][]{bld06}
\begin{equation}
\langle \dot p\rangle_{\rm p}  = {d \langle p\rangle_{\rm p} \over dt} = { 4\pi \over N_p} \int p^3 {\partial f \over \partial t} dp = 
 {4\pi \over N_p} \int p {\partial \over\partial p}\left(p^2 D_p {\partial f\over\partial p}\right),
\end{equation}
where we use Equation (\ref{eq:diffusion}) and ignore the terms $\dot f_{\rm inj}$ and $f(p)/t_{\rm esc}$. Integrating this equation by part twice, we obtain
\begin{equation}
 \langle \dot p\rangle_{\rm p}  = {4\pi \over N_p} \int f {\partial \over\partial p}\left(p^2 D_p \right). 
\end{equation}
Writing $D_p=D_0 (p/p_0)^{q}$, $dN/dp=4\pi p^2 f$, and $p/p_0 = \xi$, we obtain
\begin{equation}
 {D_0 \over p_0^2} = {N_p \langle \dot p\rangle_{\rm p}  \over 2+q}\left( \int d\xi {dN \over d\xi} \xi^{q-1}\right). \label{eq:D_0}
\end{equation}
We compute $D_0$ taking the time average of this equation all over the calculation time. We tabulate the value of $D_0$ and $A$ in Table \ref{tab:models}, where $D_0$ is normalized by $\overline{D_p}=p_0^2/t_{\rm gyro,0}$. Since $\overline{D_p}/D_0 = t_{\rm accel}/t_{\rm gyro,0}$, where we define $t_{\rm accel} = p_0^2/D_0$, we find that acceleration time is about $10^3-10^4$ times longer than the gyro period in our simulations. 

\begin{figure}[tbp]
  \begin{center}
        \includegraphics[width=8cm]{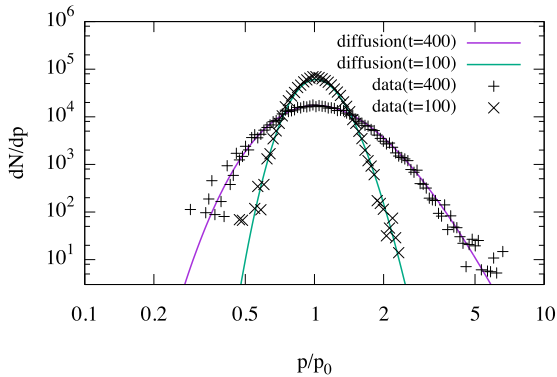}
   \caption{The evolution of the momentum distribution functions obtained by the particle simulation (points) and the diffusion equation (solid lines) for model A1. The results of the particle simulation are well described by the diffusion equation in momentum space.}
      \label{fig:dist_evolv}
    \end{center}
  \end{figure}

We solve Equation (\ref{eq:diffusion}) with $D_p$ obtained by the above procedure, and compare the distribution functions calculated by Equation (\ref{eq:diffusion}) to those obtained by the particle simulations. In Figure \ref{fig:dist_evolv}, we plot $f(p)$ of $t=100~ t_{\rm gyro,0}$ and $t=400~ t_{\rm gyro,0}$ calculated by both the simulation and the diffusion equation for model A1. We find that the distribution functions of the diffusion equation are always in agreement with those of the particle simulation. We confirm that the diffusion equation reproduces the evolution of the distribution function in all the models. We can conclude that the stochastic acceleration inside accretion flows can be described by Equation (\ref{eq:diffusion}). 

We also calculate the orbits of CRs with snapshots in the different time, $T=15~ T_{\rm rot}$ (model B1) and $T=25~ T_{\rm rot}$ (model C1). The differences of the values in Table \ref{tab:models} among models A1, B1, and C1 are less than 15\%, which indicates that our results are almost unchanged by the temporal fluctuation of the turbulent fields in our current setup of the particle simulations. However, we should note that $t_{\rm gyro,0}\simeq 6.36\times 10^{-3}~ T_{\rm rot}$ in model A1, and then the calculation time of CR orbits is longer than the time step of the MHD simulation. Thus in reality, CRs should move in a dynamically evolving turbulence. Although the current simulations with the snapshots of different times give almost the same results, it is unclear how much the dynamically evolving fields affect the orbits of CRs. This effect should be investigated as the next step in a separate paper.  

Finally, we briefly mention the scales of turbulence interacting with CRs. We show the initial gyro scales of CRs as the vertical dashed and dotted lines for model A1 and A3, respectively, in Figure \ref{fig:pwsp}. Except for model A3, the initial gyro scales are in the dissipation scale of the MHD simulation. Although the turbulence in the dissipation scale that has the steep spectra is artificially realized, the stochastic acceleration of CRs through the interaction with such turbulence is physical. Accelerated CRs can interact with the turbulence in the inertial range owing to their larger gyro scales. However, highly accelerated CRs suffer from the artificial acceleration due to the Lorentz transformation (see Subsection \ref{sec:runaway}). Thus, it is difficult to keep a large fraction of CRs in the inertial range. We need MHD data with much higher resolution to see the CRs that keep interacting with the realistic turbulent fields.

\section{DISCUSSION}\label{sec:discussion}

\subsection{Runaway Particles}\label{sec:runaway}

As briefly discussed in Section \ref{sec:numerical}, sufficiently energetic CRs in the shearing box simulation become runaway particles. When CRs of $p_y > 0$ ($p_y < 0$) cross the $x=+L_x/2$ ($x=-L_x/2$) boundary, their energies in the box frame increase. If their gyro radii are large enough to cross several boxes without changing the direction of motion, the CRs continuously gain energies.

\begin{figure}[tbp]
  \begin{center}
        \includegraphics[width=8cm]{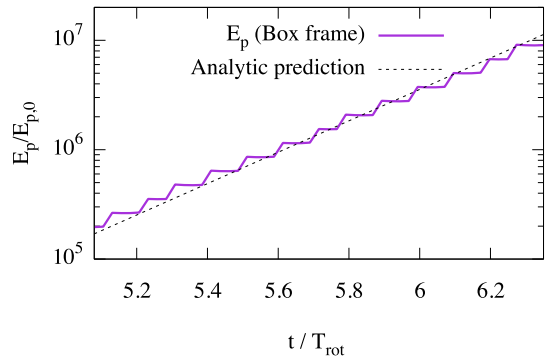}
   \caption{The temporal evolution of the energy of a runaway particle in the box frame (solid line), which is compared to the analytic estimate (dotted line). The energy of the runaway particle grows exponentially, and the energy gain rate is consistent with the analytic estimate. 
   }
      \label{fig:runaway}
    \end{center}
  \end{figure}

To see the behavior of the runaway particles, we perform a simulation without the escape boundary. Figure \ref{fig:runaway} shows the temporal evolution of energy for a runaway particle in the box frame. It is seen that the energy of the runaway particle increases stepwise every time it crosses the box boundaries in the $x$ direction. We find that the runaway particle has an almost constant box crossing time, $\Delta t = L_x/v_x\simeq L_x/(\beta_x c)$. The energy gain per crossing boundary is $\Delta E_p = \Gamma(E_p + \beta_{\rm Box} c p_y) - E_p = \left\{\Gamma (1+\beta_{\rm box}\beta_y)-1\right\} E_p$.  Thus, 
\begin{equation}
{dE_p\over dt} \approx {\Delta E_p\over\Delta t} \simeq  \left\{\Gamma (1+\beta_{\rm box}\beta_y)-1\right\}\beta_x {E_p c\over L_x }.
\end{equation}
Solving this equation, we obtain $E_p \propto \exp(t/t_{\rm grow})$, where $t_{\rm grow}\simeq  \left\{\Gamma (1+\beta_{\rm box}\beta_y)-1\right\}\beta_x L_x/c\simeq  0.30 T_{\rm rot}$, where we use $\beta_{\rm box}\simeq0.306$, $\beta_x\simeq 0.31c$,  and $\beta_y\simeq 0.92 c $, obtained by the simulation result. This solution is consistent with the temporal evolution of the energy of the runaway particle, as shown in Figure \ref{fig:runaway}. One should care about such runaway particles when using the shearing box approximation without the escape boundaries. More realistic global simulation without shearing box approximation does not suffer from this kind of unphysical effect. Particle simulations with global turbulent fields are desirable to obtain a solid conclusion, although it is computationally too expensive to perform with the required resolution. 

\subsection{Physical Interpretation of the Diffusion Coefficient}\label{sec:dp}


In this subsection, we discuss the physical interpretation of $D_p$ obtained in our simulations. As seen below, we find that the shear acceleration works efficiently, although the origin of $D_p$ is not fully understood.

The shear acceleration is expected to be efficient in accretion flows \citep[e.g.,][]{Kat91a,SBK99a}. The shear motion seems important in our simulation because the electric field is mainly produced by the shear motion. To understand its relevance clearly, we perform particle simulations assuming $V_{\rm shear}=0$ and $\beta_{\rm box}=0$. In this treatment, we do not deal with the Lorentz transformation for the orbital evolutions of CRs (equations \ref{eq:lt_e} and \ref{eq:lt_p}), and the electric field induced by shear motion vanishes. Since the runaway particle does not appear owing to $\beta_{\rm box}=0$, we set $C_{\rm esc}=\infty$. As a result, we find that the evolution of the distribution function is described by Equation (\ref{eq:diffusion}), and $D_p$ and $D_{x_i}$ is obtained in the same manner as in Section \ref{sec:results}. The resultant values are tabulated in Table \ref{tab:models} as model group X. We can see that both $q$ and $D_0$ for group X are lower than those for group A. The differences of $q$ and $D_0$ between group X and A are larger as the initial gyro radius is larger. This implies that the shear acceleration works efficiently for the particles with larger gyro radii and that it is a dominant process of particle acceleration in model A3. 

For the shearing box, $D_p$ by the shear acceleration is analytically estimated as \citep[cf.,][]{EJM88a,RD04a,Ohi13a}
\begin{equation}
 D_{p,\rm shear}\approx {1\over 15} \left({\partial V_y\over \partial x}\right)^2 \tau p^2 = {3\over 20}p^2 \tau  \Omega_{\rm K}^2, \label{eq:dpshear0}
\end{equation}
where $\tau $ is the collision time. Since we found that the diffusion coefficient in configuration space is proportional to the Bohm coefficient, we may assume 
\begin{equation}
 \tau \approx {2D_x\over v_x^2} = {\eta\over \pi} \left({p\over p_0}\right) t_{\rm gyro,0}
\end{equation}
where $\eta=D_x/D_{\rm Bohm}$ and $v_x^2=c^2/3$. Then, we can write 
\begin{equation}
 D_{p,\rm shear}\approx {3 \pi\eta\over 5} \left(t_{\rm gyro,0}\over T_{\rm rot}\right)^2 \left(p\over p_0\right)^3 \overline{D_p}.\label{eq:dpshear}
\end{equation}
We find $D_{p,\rm shear}\propto p^3$ and $D_{p,\rm shear}/\overline{D_p}\simeq 6.1\times 10^{-4}$ for $p=p_0$, using the parameter set of model A3 ($\eta\simeq2.0$ and $t_{\rm gyro,0}/T_{\rm rot}\simeq1.3\times10^{-2}$). The momentum dependence and value of this estimate are almost consistent with $D_p$ of the simulation result, which also shows that the shear acceleration is dominant process for model A3. 


Then a remaining problem is the physical origin of $D_p$ in model X, in which CRs are accelerated by the electric field induced only by the turbulent fluid motion with the velocity $\vect {\delta V}$. According to the quasi-linear theory of gyro-resonance by isotropic Alfv\'{e}n modes, the power index of $D_p$ is the same as the index of the power spectrum of the turbulent magnetic field \citep[e.g.,][]{sp08}, in which $q=5/3$ is expected for the Kolmogorov turbulence. For our turbulence, the power spectra are steeper for smaller scales (see Figure \ref{fig:pwsp}), so this theory would predict higher values of $q$ for lower $p_0$ models. Model group A shows the opposite trend, while model group X is qualitatively consistent with this feature. However, quantitatively, $q\sim4$ for model X1 (large $k$ due to low $E_{p,0}$) and $q\sim5/3$ for model X3 (small $k$ due to high $E_{p,0}$) are expected in this theory, which are far from the particle simulation results.

Our turbulence has the anisotropic power spectra (see Figure \ref{fig:pwsp}). \citet{YL02a} investigate the scattering frequency of such anisotropic MHD turbulences and show that the scattering frequency by Alfv\'{e}n modes is sufficiently suppressed, compared to the case of isotropic turbulences. Instead, gyro-resonance by fast modes plays a crucial role in scattering CRs under the MHD turbulences. The fast modes are almost isotropic, and the index of its power spectrum are 1.5 \citep{CL03a}, which leads to $q=1.5$ \cite[cf.][]{CL06a}. Model X2 has the value close to $q=1.5$, but it is unlikely that the fast modes have substantial power in the MRI turbulence because velocity fluctuations are sub-sonic in our turbulent field. 

\citet{lyn+14} claim that for a sub-sonic MHD turbulence, the fast modes have only a small fraction of the turbulent power and slow modes dominate over the other modes in acceleration of relativistic CRs. They consider the mirror force acting on CRs, and analytically derive $D_p \propto p^2$ for stochastic acceleration in the weakly compressible turbulence. In this case, $q$ should always be equal to 2, which is often called the hard-sphere model, but this theory does not seem to be applicable to our model since the resultant $q$ of our simulations is not close to 2 for both group X and A. 



\subsection{Plasma Properties in RIAFs}\label{sec:plasma}

As the first step study, we use the shearing box approximation for RIAFs, which is also used in the previous studies \citep[e.g.][]{SHQ06a,hos13}. However, the shearing box approximation has a few points at odds with RIAFs. First, the local approximation is not good for RIAFs because the aspect ratio, $H/r$, is close to unity because of the high temperature. The centrifugal balance is also violated due to the strong pressure gradient force in RIAFs. The angular velocity is usually sub-Keplerian, which seems to weaken the shear acceleration. In addition, the radial advection of heat is important for the global structure of RIAFs. The high temperature makes viscous stress strong under the $\alpha$ prescription, which makes radial velocity fast, compared to the standard disk \citep[e.g.,][]{ny94,ktt14}. The global simulations can take account of these effects consistently. 

The magnetic reconnection is supposed to play an important role in the saturation of MRI turbulence \citep[e.g.,][]{SI01a}. Recent PIC simulations also suggest that the dissipation by reconnection of stretched magnetic fields is important for both the saturation of MRI turbulence and generation of CRs \citep{hos15}. However, since we solve a set of ideal MHD equations in which magnetic reconnection happens due to the numerical resistivity, we ignore the electric field generated by finite resistivity in the simulation of CRs. In reality, the electric field generated at reconnection site is expected to enhance the particle acceleration in MRI turbulence. To include this effect, the MHD simulation that includes the resistivity model based on collision-less plasma physics is required, which is beyond the scope of the present paper. 

The effect of back reaction by CRs on the non-linear evolution for MRI is not well understood. \citet{KK15a} analyze the linear growth of MRI including the effect of CR diffusion. They show that the CR diffusion barely affects the linear growth of MRI in shearing box calculation, even if the CR pressure is comparable to the thermal pressure. However, the effect of CRs on the saturation level of MRI turbulence has not been studied yet.
In the case close to ideal MHD, the saturation level is supposed to be determined by the parasitic instability of Kelvin-Helmholtz (KH) modes that induce the dissipation by magnetic reconnection
 \citep{GX94a,SI01a,san+04,Pes10a}. Since the growth rate of KH instability is modified by CRs if they have comparable energy density to the thermal particles \citep{STK14a}, CRs probably affect the saturation level of MRI turbulence. To investigate it, three dimensional MHD simulations including the back reaction by CRs are necessary \citep[cf.,][]{BCS15a}.

Although we assume that the thermal component is described by the MHD formalism, it is unclear whether the collisionless plasma is described by MHD. As an alternative way to describe the collisionless plasma in accretion flows, \citet{QDH02a} used the anisotropic MHD formalism where the plasma has anisotropic pressure. In this formalism, as the magnetic field becomes strong, the pressure anisotropy increases, which suppresses the growth of MRI \citep{SHQ06a}. In reality, the thermal particles are expected to be isotropized by some plasma instabilities such as the mirror mode, so that the pressure anisotropy decreases. This allows MRI to grow up in non-linear regime, which is confirmed by the recent PIC simulations with the shearing box approximation \citep{riq+12,hos13,hos15}. However, as noted in Section \ref{sec:intro}, there is the scale gap between the gyro scales of electrons and the scale heights of accretion flows. The hybrid simulations, in which we treat protons as particles and electrons as a fluid, may be useful for investigating larger scale turbulences with the kinetic effects \cite[cf.,][]{SH14a}, but it is currently difficult to bridge the scale gap completely.



\section{SUMMARY}
\label{sec:summary}

We have investigated the stochastic particle acceleration process in accretion flows. The CRs are likely to exist inside accretion flows when the mass accretion rate is sufficiently lower than the Eddington rate \citep[e.g.][]{ktt14}. The existence of CRs is expected by the theoretical estimate \citep[e.g.,][]{kmt15}, numerical simulations \citep[e.g.,][]{hos15}, and modeling of the observed photon spectra \citep[e.g.,][]{nse14}, although the gamma-rays from accretion flows are not detected yet \citep{WNX15a}.

We have calculated the orbits of CRs as test particles in turbulent fields generated by MRI with the shearing box approximation. Within this approximation, when a CR particle crosses the box boundary in the $x$ direction, it changes energy as measured in the box frame. Sufficiently energetic CRs that can cross the box boundary in the x direction many times without changing the direction of motion tend to increase their energies in a runaway manner. We have avoided this artificial acceleration by introducing the escape boundaries: The CRs crossing them are allowed to escape from the system. From our simulations, we have found that the energies of CRs randomly change. Some lucky CRs continuously increase their energies, and reach several times higher energies than their initial energies after $t\sim400t_{\rm gyro,0}$. The CRs anisotropically diffuse in the configuration space. Since the magnetic field is stretched mainly in the $y$ direction, the diffusion coefficients satisfy $D_y > D_x \sim D_z$. The energy dependence of $D_{x_i}$ is consistent with the Bohm coefficient, and we can approximately write $D_y\sim 20 D_{\rm Bohm}$ and $D_x\sim D_z\sim 2 D_{\rm Bohm}$. 

We have also shown that the evolution of distribution function is described by the diffusion equation in momentum space (Equation \ref{eq:diffusion}). The diffusion coefficient in momentum space can be fitted by a power-law form, $D_p=D_0(p/p_0)^q$. Both $D_0$ and $q$ depend on the initial momentum $p_0$. Although the shear acceleration is efficient for sufficiently energetic particles, quantitative understanding of the acceleration mechanism in our simulation is not complete. To apply our simulation results to astrophysical objects, e.g. calculating the spectrum of non-thermal protons and predict the neutrino and gamma-ray fluxes more precisely than those in \citet{kmt15}, we need to understand the acceleration process and the dependence of $D_p$ on physical quantities.




\acknowledgments
We thank Masahiro Hoshino, Jun Kakuwa, Yutaka Ohira, Ryo Yamazaki, Daisuke Nakauchi, and Sanemichi Takahashi for fruitful discussion. This work is partly supported by JST grant ``Building of Consortia for the Development of Human Resources in Science and Technology'' (S.S.K. and K.T.) and JSPS Grants-in-Aid for Scientific Research 15H05437 (K.T.).

\clearpage


\end{document}